\documentclass[journal]{IEEEtai}

\ifCLASSINFOpdf
\else
\fi

\usepackage{url}
\usepackage[colorlinks,linkcolor=black, anchorcolor=black, citecolor=black,urlcolor=black]{hyperref}
\usepackage{caption}
\usepackage{stmaryrd}
\usepackage{amsfonts}
\usepackage{graphicx}
\usepackage{amsmath,bm}
\usepackage{multirow,makecell}
\usepackage{caption}
\usepackage{color}
\usepackage{graphicx,times,amsmath} 
\usepackage{graphicx}
\usepackage{amsmath,mathrsfs}
\usepackage{amssymb}
\usepackage{booktabs}
\usepackage{stfloats}
\usepackage{subfigure}
\usepackage[ruled,vlined]{algorithm2e}

\SetCommentSty{mycommfont}

\makeatletter
\newcommand{\algorithmfootnote}[2][\footnotesize]{%
  \let\old@algocf@finish\@algocf@finish
  \def\@algocf@finish{\old@algocf@finish
    \leavevmode\rlap{\begin{minipage}{\linewidth}
    #1#2
    \end{minipage}}%
  }%
}
\makeatother

\usepackage{threeparttable}
\usepackage[numbers,sort&compress]{natbib}

\definecolor{orange}{rgb}{1,0.4,0}

\begin{document}

\title{\ \\ \LARGE\bf Multi-modal Learning based Prediction for  Disease}

\author{Yaran Chen$^{\dagger}$, Xueyu Chen$^{\dagger}$, Yu Han$^{\dagger}$, Haoran Li, Dongbin Zhao$^{*}$~\IEEEmembership{Fellow,~IEEE}, Jingzhong Li and Xu Wang$^{*}$
\thanks{ Y. Chen Y. Han, H. Li and D. Zhao are with the State Key Laboratory of Multimodal Artificial Intelligence Systems, Institute of Automation, Chinese Academy of Sciences, Beijing 100190, and also with the College of Artificial Intelligence, University of Chinese Academy of Sciences, Beijing 100049, China. (email: chenyaran2013@ia.ac.cn, hanyu2021@ia.ac.cn, lihaoran2015@ia.ac.cn, dongbin.zhao@ia.ac.cn) }
\thanks{ X. Chen is with the Department of Biostatistics, School of Public Health, Cheeloo College of Medicine, Shandong University, Jinan, China. (email: chenxueyv@163.com) }
\thanks{ J. Li is with Information and Educational Technology Center, Beijing University of Chinese Medicine, Beijing, 100029, China. (email: lijz@bucm.edu.cn)}
\thanks{ X. Wang is with School of Life Sciences, Beijing University of Chinese Medicine, Beijing, 100029, China. (email: wangx@bucm.edu.cn) }
\thanks{This work is partly supported by the National Natural Science Foundation of China (NSFC) under Grants No. 62173324 and No.82004222, the National Key R\&D Program of China No. 2017YFC1700106, and the Key Research Program of the Chinese Academy of Sciences No. ZDRW-ZS-2021-1-2.}
\thanks{$^{*}$: D. Zhao and X. Wang are the corresponding authors.}
\thanks{ $^{\dagger}$: They contribute equally to this work.}
}


\markboth{Journal of \LaTeX\ Class Files,~Vol.~14, No.~8, August~2015}%
{Shell \MakeLowercase{\textit{et al.}}: Bare Demo of IEEEtran.cls for IEEE Journals}


\maketitle
\IEEEpeerreviewmaketitle


\begin{abstract}
   Non-alcoholic fatty liver disease (NAFLD) is the most common cause of chronic liver disease, which can be predicted accurately to prevent advanced fibrosis and cirrhosis. While, a liver biopsy, the gold standard for NAFLD diagnosis, is invasive, expensive, and prone to sampling errors. Therefore, non-invasive studies are extremely promising, yet they are still in their infancy due to the lack of comprehensive research data and intelligent methods for multi-modal data. This paper proposes a NAFLD diagnosis system (DeepFLDDiag) combining a comprehensive clinical dataset (FLDData) and a multi-modal learning-based NAFLD prediction method (DeepFLD). 
   The dataset includes over 6000 participants' physical examinations, laboratory and imaging studies, extensive questionnaires, and facial images of partial participants, which is comprehensive and valuable for clinical studies. From the dataset, we quantitatively analyze and select clinical metadata that most contribute to NAFLD prediction.  Furthermore, the proposed DeepFLD, a deep neural network model designed to predict NAFLD using multi-modal input, including metadata and facial images, outperforms the approach that only uses metadata.   Satisfactory performance is also verified on other unseen datasets. Inspiringly, DeepFLD can achieve competitive results using only facial images as input rather than metadata, paving the way for a more robust and simpler non-invasive NAFLD diagnosis. 

 \end{abstract}
 \begin{IEEEImpStatement}
 Non-alcoholic fatty liver disease  is the leading cause of chronic liver disease worldwide and its prevalence is increasing. Current studies usually use methods such as blood and metabolic indicators, which are only accurate up to 80\%, leading to multiple missed and false detections, and also delaying the patient's condition. By building a database combining face and physical examination data, we propose a fatty liver detection method that merges face and indicator multimodal data, increasing detection accuracy to 87.5\%. The proposed method also allows the use of the face to assess the presence of fatty liver, providing simple and rapid diagnosis in rural areas with limited access to medical care.
\end{IEEEImpStatement}


\section{Introduction}

\IEEEPARstart{C}{hronic} liver diseases are common causes of morbidity and mortality worldwide, and liver-related diseases account for over 2 million deaths per year worldwide \cite{RN3}. Non-alcoholic fatty liver disease (NAFLD), also  known as metabolic-associated fatty liver disease, is one of the most common chronic diseases and metabolic complications of obesity \cite{RN4}. As obesity rapidly increases, the prevalence of NAFLD is increasing globally, ranging from approximately 30\% in the general population to approximately 80\% in morbidly obese individuals \cite{RN2}. NAFLD, a spectrum of liver abnormalities ranging from NAFLD to non-alcoholic steatohepatitis (NASH), is predicted to be the most common indication for liver transplantation by 2030 \cite{RN23, RN6}. NAFLD is characterized by excessive fat accumulation and is the major risk factor for the development of NASH, liver fibrosis, and cirrhosis \cite{RN33}. 
Early diagnosis and treatment are critical to reducing associated complications and mortality.

For centuries, physicians in the clinic have diagnosed NAFLD by several methods, of which liver biopsy has been evaluated as the gold standard, yet it is featured as invasive  and expensive \cite{RN7}. Some radiological techniques and ultrasonography are effective alternatives for liver biopsy, but they have  limited access to remote areas due to the high cost of  instruments and tests \cite{RN9}. Therefore, non-invasive and inexpensive NAFLD diagnosis methods have been extremely promising.


There have been several alternative methods for detecting NAFLD in previous studies. \cite{RN35} used a machine learning model to classify NAFLD according to human serum and stool. In addition, FIB-4, NFS, and neck circumference have also been studied to diagnose  NAFLD \cite{RN39, RN38}. However, most of the above methods are concentrated only on unilateral factors, which  may be due to the lack of comprehensive data. Considering multifaceted perspectives for NAFLD prediction could be effective.

In order to consider NAFLD prediction from multiple perspectives and multiple views, We intend to create a comprehensive clinical database that includes questionnaires, physical examinations, laboratory tests, and imaging examinations (blood routine examination, urinalysis, and so on). In fact, the image of the face is a convenient window into the internal organs' function. Facial images have been used as an important diagnostic tool in traditional Chinese medicine and Western medicine clinical fields\cite{RN10, RN11}. At present, studies have proven that human facial features can reflect developmental syndromes, biological age, and the aging degree of organs\cite{RN12, RN13}. Many studies have fully proven the auxiliary value of facial images in disease diagnosis, which can easily and conveniently help clinicians make disease judgments, especially in traditional Chinese medicine. Recently, deep convolutional neural networks (CNNs), as one of the most efficient networks in computer vision \cite{D.Zhao,Y.Chen,DBLP:journals/tcyb/ChenGLZ22}, have been widely used for image-based disease diagnoses, such as COVID-19, heart disease and small-bowel disease\cite{9664332, 9440796, RN15, 2020Three}. CNN-based deep-learning algorithms have achieved near-human-level performance in disease classification, and even surpassed humans in subtle points that humans cannot observe\cite{RN16}. Therefore, facial images, which can be acquired rapidly, non-invasively, and freely, may be  potential and essential information for the screening and prediction of NAFLD.  In this study, we aim to build comprehensive clinical data, including facial images, physical examinations, and so on. To the best of our knowledge, no prior studies have exploited the association between facial images and NAFLD.



In this paper,  a NAFLD diagnosis system is developed to distinguish NAFLD using multi-modal input, including facial images and metadata. First, we collect volunteers' physical examinations, laboratory and imaging studies, questionnaires, and facial images to construct a medical dataset FLDData. Then we adopt a joint indicator-based data analysis method  to quantitatively analyze and select the clinical metadata most relevant to NAFLD from the medical dataset. Based on the selected data, we propose a multi-modal based NAFLD prediction method DeepFLD, including the facial images and metadata. Since NAFLD is complex, it is difficult to extract effective features directly from facial images.
In DeepFLD, a medical constraints-based auxiliary task is designed to extract valid image features. 
Compared with the indicators selected by considering only the Pearson Correlation Coefficient, the indicators we selected can improve the classification accuracy of NAFLD. The proposed DeepFLD with multi-modal as input outperforms the models with metadata as input and achieves acceptable performance on unseen data. DeepFLD can achieve competitive results using only facial images as input rather than metadata, which is encouraging.

To sum up, our contributions in this paper are mainly as follows:
\begin{itemize}
   \item A multi-modal human clinical dataset is built by collecting facial images, physical examination information, laboratory examination and imaging studies, and questionnaires. We also quantitatively analyze  the most relevant medical indicators for NAFLD using a joint indicator-based data analysis method.  
   
   \item 
   
   We propose a NAFLD prediction method DeepFLD with multi-modal input and medical constraints, that facilitates valid feature extraction from the high-dimensional facial image. As the first to introduce facial images into NAFLD prediction, DeepFLD with multi-modal input outperforms other methods with only metadata as input, and verifies satisfactory performance on other unseen datasets. Furthermore, when compared to metadata, DeepFLD can achieve competitive results with only facial images as input, providing a viable way for a more robust and simpler non-invasive diagnosis of NAFLD.
   
   \item We analyze the NAFLD prediction results by exploring the facial characteristics of the person who has NAFLD. Among these characteristics, dark skin color and the presence of melasma can be supported by previous medical studies. 
\end{itemize}

The rest of this paper is organized as follows: In Sec. II, we give a review of the related work; Sec. III introduces the prediction system, including the collected dataset and the prediction models; In Sec. IV, we present our experiments and results from the collected dataset and analyze the lifestyle; Finally, we conclude the study in Sec. V.


\section{Related Work}

For decades, there have been many ways to detect NAFLD. Liver biopsy has been evaluated as the gold standard for diagnosing the presence and extent of liver inflammation and fibrosis but has the disadvantage of being invasive\cite{RN7}. Several radiological techniques and ultrasonography are effective alternatives to liver biopsy for quantification of hepatic steatosis (Computed Tomography, Magnetic Resonance Imaging, or Magnetic Resonance Spectroscopy) and fibrosis (Transient Elastography), but screening of fatty liver populations is more problematic due to expensive or limited access to remote areas\cite{RN9, RN8}. Invasiveness, high cost, and inconvenience hinder their detection in large populations and screening in remote populations.

Several alternative methods for detecting NAFLD have been identified in previous studies, but the search for alternatives for sensitive NAFLD detection is still ongoing.  \cite {RN35} explores human serum and feces for metabolomics studies to predict NAFLD. It used a machine learning method to predict NAFLD in 180 participants with a final AUC value of 0.72-0.80. Furthermore, several studies of alternative diagnostic methods have investigated the expression status of genes involved in cholesterol, triglyceride, and other lipid metabolism and identified NAFLD-related genes \cite{RN37}. \cite{RN36} shows that obesity significantly amplifies the effects of three gene sequence variants associated with NAFLD. The synergistic effect between adiposity and genotype promotes NAFLD pathology, from steatosis to liver inflammation to cirrhosis.
At the same time, there are also studies on the diagnosis and prediction of NAFLD by scoring. \cite{RN39} explores the effect of neck circumference on NAFLD by recruiting 674 patients. Neck circumference was found to be associated with fatty liver prevalence, but it could not predict NAFLD on its own. By recruiting 5129 volunteers and measuring their (FIB-4) and (NFS), \cite{RN38} explores whether FIB-4 and NFS could be used to screen for NAFLD. While the results are not satisfactory. However, although the index was convenient, it could not accurately evaluate and screen for disease. Diagnosis by genomics or metabolomics is relatively reliable but difficult for large-scale screening and monitoring in community-based populations.

\section{The DeepFLDDiag System}
\begin{figure*}[htb!]
   \vspace{-0.1cm}
   \centerline{
      \includegraphics[width=0.9\linewidth]{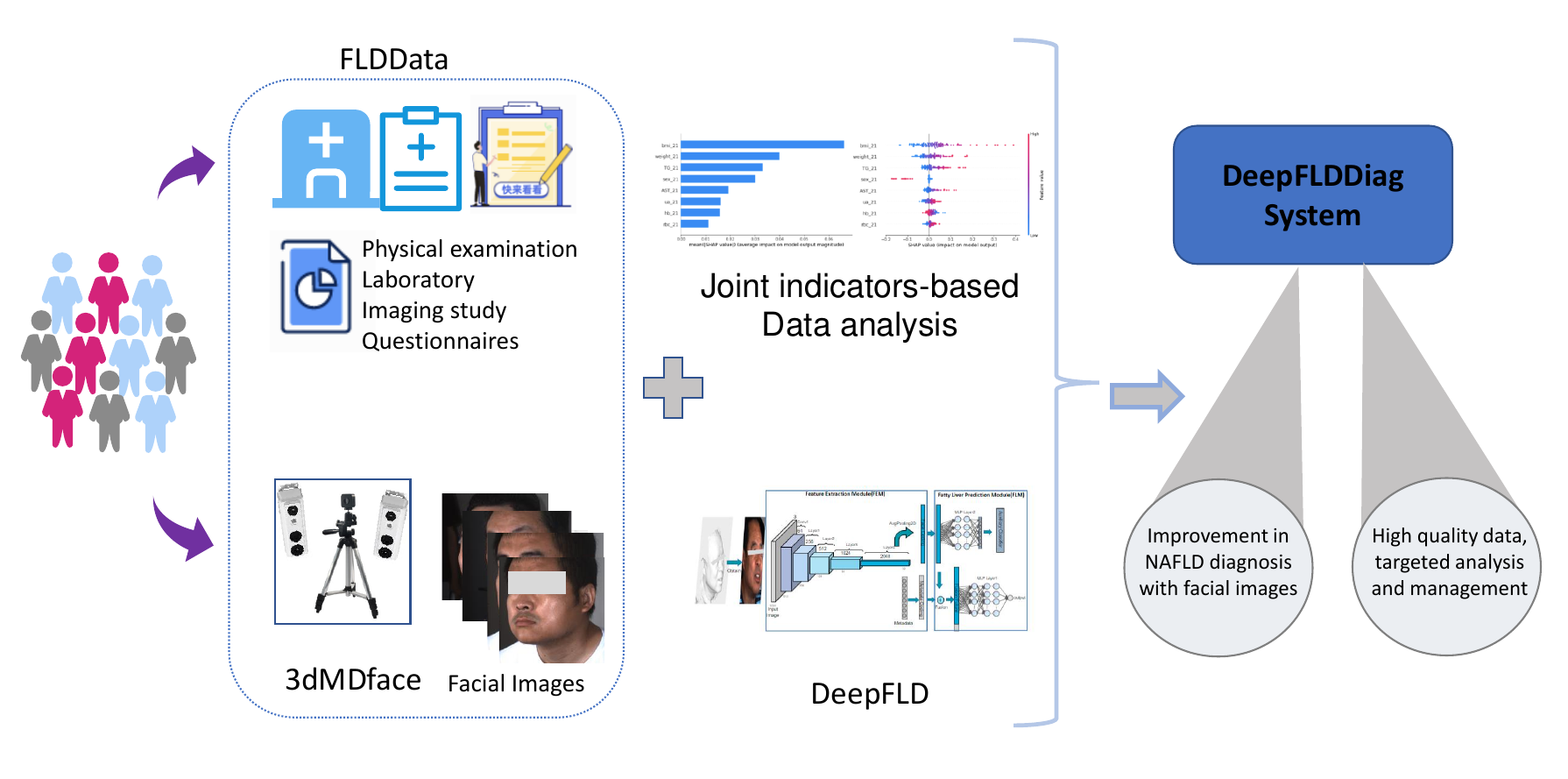}}
   \caption{The AI system DeepFLDDiag for NAFLD prediction. We capture volunteers’ facial image data using the 3dMDface instrument and collect their medical metadata from their physical examinations, extensive questionnaires, and so on. Through the proposed data analysis and machine learning method, we can improve in NAFLD diagnosis with facial images and high quality data, targeted analysis and management.}
   \label{fig:uni-view}
   \vspace{-0.1cm}
\end{figure*}

In this section, we present the proposed NAFLD diagnosis system DeepFLDDiag, which consists of three parts: data collection, data processing and disease prediction, as shown in Fig. \ref{fig:uni-view}. First, we build the dataset FLDData, including the data of facial images. Second, a method of data analysis is adopted to quantitatively analyze and select the most correlated clinical metadata with NAFLD from medical datasets. Finally, we design a  NAFLD prediction method called DeepFLD. In particular, it is important to be reminded that: 1) The data we collected contains very comprehensive medical information about volunteers: face data and data of 480-dimensional indicators, including physical examination data and questionnaire data; 2) In the data analysis, we consider not only the influence of individual factors on the data, but also the use of a joint indicators method to mine the impact on the results in the case of multiple factors combined; 3) In DeepFLD, we designed an auxiliary task based on multi-modal data fusion and, medical constraints to extract valid image features.
The following sections specifically describe each component.

\subsection{FLDData}
FLDData is compiled by gathering information from over 6,000 volunteers who take part in the Jidong Cohort Study (COACS) in Tangshan City (Hebei Province, North China). The study is an observational, prospective, community-based cohort dataset. All participants underwent a physical examination, laboratory and imaging studies, extensive questionnaires, and facial image acquisition. The research ethics approval for this study was obtained from the Medical Ethics Committee, Staff Hospital, Jidong Oilfield Branch, and China National Petroleum Corporation, and followed the recommendations of the Declaration of Helsinki. All participants originally  signed a written informed consent form.

\subsubsection{Data acquisition}
During the physical examination, weight, height, waist and hip circumferences, and blood pressure are measured. Blood samples for laboratory examinations are collected after an overnight fast. Imaging results included ultrasonography and computed tomography. Meanwhile, information on medical history, medication use, lifestyle, and other sociodemographic factors was obtained by interview and questionnaire. Including the information mentioned above, each participant's 480-dimensional metadata is recorded in  FLDData.


To explore the relationship between facial images and NAFLD, we use a camera to collect the participants' corresponding RGB and depth images and synthesize images. Facial image acquisition and quality control are made in accordance with the protocol \cite{RN101}. Participants set up in designated positions and look at the camera for 1 minute, with no face covering, for the facial images to be taken. 

 In order to ensure the balance of the data distribution, the collected data is as balanced as possible.
 Overall, 6760 participants are enrolled in the cross-sectional analysis, and 49.9\% of them have NAFLD, of whom 68.9\% are male. Among participants, the mean age is 44 years old.

\renewcommand\arraystretch{1.2} 
\begin{table*}
   \footnotesize  
   \begin{center}
   \caption{Distribution and ranking of some indicators according to Pearson Correlation Coefficient}
   \begin{threeparttable}
   \begin{tabular}{cccccc}
   \hline
   \hline
    Variables & Total 	&NAFLD&	NON-NAFLD&	$|\rho|^*$  & Note\\
   \hline
   BMI ($\rm{kg/m^2})\dagger$ &	24.4$\pm$3.7&	26.3$\pm$3.2	&22.0 $\pm$2.4 	&0.602 &a	Body Mass Index\\
   WEIGHT ($\rm{kg}$)\tnote{$\dagger$}  &	67.3 $\pm$ 16.9&	76.0 $\pm$ 12.8	&61.0 $\pm$ 9.3	& 0.547 & Weight\\
   HLP n (\%)	&3024 (46.3)	&2346 (63.5)	&678 (24.0)	&0.392	&Hyperlipidaemia\\
   UA ($\rm{umol/L})\dagger$&	354.0 $\pm$ 95.6	&385.2 $\pm$ 95.2	&312.0 $\pm$ 78.1 &0.381	&Uric Acid\\
   TG ($\rm{mmol/L})\dagger$ &	1.8$\pm$  1.6	&2.32 $\pm$  1.9	&1.18 $\pm$  0.6 	&0.362	&Triglyceride \\
   OBE n (\%)&	942 (15.4)	&911 (26.3)&	31 (1.2)	&0.346	&Obesity\\
   DBP ($\rm{mmHg})\dagger$	&75.2 $\pm$  12.3&	78.5 ± 12.3&	70.3 $\pm$  10.4&0.332	&Diastolic  Blood Pressure\\
   SBP ($\rm{mmHg})\dagger$	&124.1 $\pm$  18.8&	129.6 $\pm$  18.8&	117.1 $\pm$  16.0	&0.329	&Systolic Blood Pressure\\
   APOB ($\rm{g/L})\dagger$	&0.9 $\pm$  0.2	&1.0 $\pm$  0.2	&0.8 $\pm$  0.2	&0.328&	Apolipoprotein B\\
   HGB($\rm{g/L})\dagger$	&140.1 $\pm$  15.9&	145.2 $\pm$  14.8	&134.8 $\pm$  15.4	&0.324	&Hemoglobin\\
   RBC ($\rm{10^12/L})\dagger$	&4.5 $\pm$  0.4	&4.6 $\pm$  0.4&	4.4 $\pm$  0.4	&0.321	&Red Blood Cell\\
   MALE n (\%)	&3351 (51.0)	&2394 (64.4)	&957 (33.5)	&0.306	&Male \\
   AST ($\rm{U/L})\dagger$	&23.6 $\pm$  11.7	&25.5 $\pm$  12.5	&20.9 $\pm$  8.8	&0.304	&Aspartate Aminotransferase\\
   HUA n (\%)	&1828 (28.0)	&1431 (38.7)	&397 (14.0)	&0.272	&Hyperuricemia\\
   HPT n (\%)	&1441 (23.0)	&1149 (32.4)	&292 (10.7)	&0.255	&Hypertension\\
   HDL $(\rm{mmol/L})\dagger$	&1.4 $\pm$  0.4	&1.3 $\pm$  0.3	&1.5 $\pm$  0.4	&0.252	&High Density Lipoprotein Cholestero \\
   WBC $(\rm{10^9/L})\dagger$	&6.6 $\pm$  1.6	&6.9 $\pm$  1.7	&6.1 $\pm$  1.5	&0.242	&White Blood Cell\\
   LDL$(\rm{mmol/L})\dagger$	&2.0 $\pm$  0.7	&2.2 $\pm$  0.7	&1.8 $\pm$  0.7	&0.238	&Low Density Lipoprotein Cholestero \\
   APOA $(\rm{g/L})$	&1.4 $\pm$  0.3	&1.3 $\pm$  0.3	&1.5 $\pm$  0.3	&0.223	&Apolipoprotein A\\
   ALP $\rm{(U/L)}\dagger$	&71.8 $\pm$  23.6	&75.7 $\pm$  20.9	&66.0 $\pm$  22.2	&0.219	&Alkaline Phosphatase\\
   FBG $\rm{(mmol/L)\dagger}$	&5.8 $\pm$  1.5	&6.0 $\pm$  1.7	&5.4 $\pm$  1.0	&0.217	&Fasting Blood Glucose\\
   \hline
   \hline
\end{tabular}
    \begin{tablenotes}
         \footnotesize
         \item[$\dagger$]  Results are presented as  mean $\pm$ s.d.
         \item[$^*$] Absolute value of Pearson Correlation Coefficient 
      \end{tablenotes}
\end{threeparttable}
\label{tab:21indicators}
   \end{center}
\end{table*}

\subsubsection{Assessment of NAFLD}
NAFLD is diagnosed based on abdominal ultrasonography (ACUSON X300, Siemens, Munich, Germany) using a 3.5-MHz probe that is performed in all participants by skilled sonographers following a standardized protocol. We assess NAFLD according to the standard criteria established by the Asia-Pacific Working Party on NAFLD and the Chinese Association for the Study of Liver Disease \cite{RN32,RN31}. After the exclusion of  subjects with excessive alcohol intake or other hepatic diseases, participants with NAFLD must present two or more of the following abnormal characteristics: (1) diffusely increased liver near-field echogenicity relative to the kidney; (2) ultrasound beam attenuation; and (3) poor visualization of intrahepatic structures.

\subsection{Joint indicator-based Data analysis}
\label{data_analysis}
In FattyLiverData, we have collected 480 indicators for each participant, including the content of physical examinations such as height and weight, and the content of questionnaires on living habits. We intend to select the most relevant data for NAFLD from these indicators because they are redundant and duplicate. 

The correlation coefficient is a basic similarity metrics between variables in statistics and is used to analyze FLDData in this paper. 
Although we can obtain the correlation  between a certain indicator and NAFLD, we cannot obtain the correlations between multiple indicators and NAFLD. It is known that NAFLD is a complex disease caused by multiple factors. So these indicators selected by the correlation coefficient do not certainly contribute to NAFLD prediction. In order to obtain effective indicators for  NAFLD prediction, we consider  the impact of multiple indicators on NAFLD. Specificity, we design a joint indicators-based data analysis method that combines correlation coefficient (Pearson Correlation Coefficient) and model analysis method (SHAP) to obtain the effect of a combination of indicators on NAFLD. Then the most helpful indicators for NAFLD prediction are selected from FLDData.


\subsubsection{Pearson Correlation Coefficient} As a common correlation coefficient, Pearson Correlation Coefficient, which ranges from -1 to 1, can measure the correlation between two variables. First, we use  Pearson Correlation Coefficient to analyze FLDData and identify the indicators that are most closely related to NAFLD.  In  mathematical form, Pearson Correlation Coefficient of two variables $(p,q)$ can be described as follows:
\begin{equation}
   \rho (p,q) = \frac{\sum{pq}-\frac{\sum{p}\sum{q}}{N}}{\sqrt{(\sum{p^2}-\frac{(\sum{p})^2}{N})(\sum{q^2}-\frac{(\sum{q})^2}{N})}}
\end{equation}
where $N$ is the number of attributes total. The symbol  $|\rho(p,q)| =1$  indicates that $p$ and $q$ are perfectly correlated, corresponding to the modified Euclidean distance. The stronger the correlation between $(p,q)$ the lower the value of $rho(p,q)|$.

Considering that many indicators in our data are very redundant for  NAFLD, we selected 21 indicators by calculating the Pearson Correlation Coefficient  coefficients between NAFLD and all other indicators, as shown in Tab. \ref{tab:21indicators}. 

We also analyze the selected 21 indicators using statistical analysis (SAS 9.4). In Tab. \ref{tab:21indicators}, continuous variables are expressed as means and standard deviations, and categorical variables are expressed as frequencies and percentages. The selected 21 indicators are compared using two-sample independent t-tests, and P$<$0.05 is set as the level of significance.
The NAFLD group's indicators have significantly higher means than the NON-NAFLD group's. Meanwhile, significant differences are observed in those indicators from NAFLD/NON-NAFLD groups by the statistical analysis (P$<$0.05).

\subsubsection{SHapley Additive exPlnations(SHAP)}
SHAP \cite{lundberg2017unified} quantifies the impact of each indicator on the results in the model and is often used in the interpretability of black box models. For the selected 21 indicators, we adopt SHAP to further analyze the contribution of each indicator to the NAFLD prediction. According to expert experience, people who smoke and drink are also prone to acquiring NAFLD. However, they are not selected by Pearson Correlation Coefficient. Therefore, we add the two indicators to the 21 selected  indicators to be considered together.
First, we build a neural network for NAFLD prediction with the 23 indicators (denoted as $Metadata_{23}$) as inputs. Then we calculate the SHAP value for each indicator, which is the contribution to the NAFLD prediction.
For the set $U\subseteq\{x_1,x_2,\cdots ,x_p\}$  of selected indicators, where $x_j$ denotes the $j-$th indicator, the SHAP value of $x_j$ is calculated as follows:
\begin{equation}
   \resizebox{.91\hsize}{!}{$\phi _j=\sum_{S\subseteq \{x_1,\cdots ,x_p\}\diagdown {x_j}} {\frac{|S|!(p-|S|-1)!}{p!} (f_x(S\bigcup {x_j})-f_x(S))}$}
   \label{eq:shap}
\end{equation}
where ${\{x_1,\cdots ,x_p\}\diagdown {x_j}}$ denotes the subset $V$ without the indicator $x_j$, 
$S$ is a subset of set $V$, and $\frac{|S|!(p-|S|-1)!}{p!}$ refers to the probability of  $S$  occurrence. The NAFLD prediction model $f_x(S)$ refers to the prediction performance when the indicator set $S$ is used as input.

Given the current set of indicators,  the estimated Shapley value is calculated based on the actual prediction performance and the average prediction performance (shown in (\ref{eq:shap})). 

\begin{figure*}[htbp]
   \centering
    \subfigure[Mean value of the SHAP value.]{
     \includegraphics[width=0.4 \linewidth]{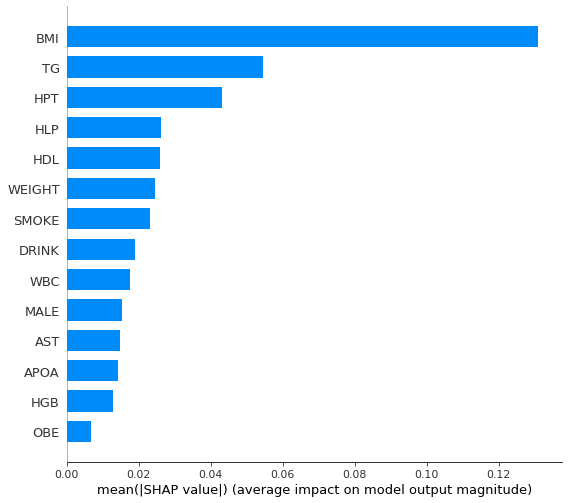}}
   \label{2a}\hfill
    \subfigure[ SHAP value.]{
       \includegraphics[width=0.4 \linewidth]{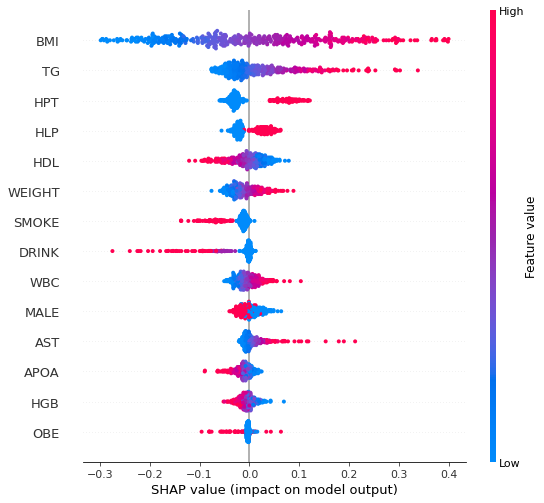}}
   \label{2b}
    \caption{Ranking of indicators according to SHAP value. The higher the SHAP value of the indicator is, the larger the contribution to NAFLD prediction is, in which BMI is the largest contributor to the prediction of NAFLD.}
    \label{fig:shap} 
\end{figure*}

By calculating the SHAP value, we obtain the indicator contribution ranking, shown in Fig. \ref{fig:shap}. It shows that BMI has the largest contribution to NAFLD prediction. Compared to the ranking of Pearson Correlation Coefficients (Tab. \ref{tab:21indicators}), the ranking of the SHAP value has changed significantly. The  two added indicators Smoke and Drink play an important role in the network.  Therefore, through the data analysis, we can choose the indicators with the largest effectiveness.

\subsection{DeepFLD Prediction Method}

\begin{figure*}[htb!]
   \vspace{-0.1cm}
   \centerline{
      \includegraphics[width=0.9\linewidth]{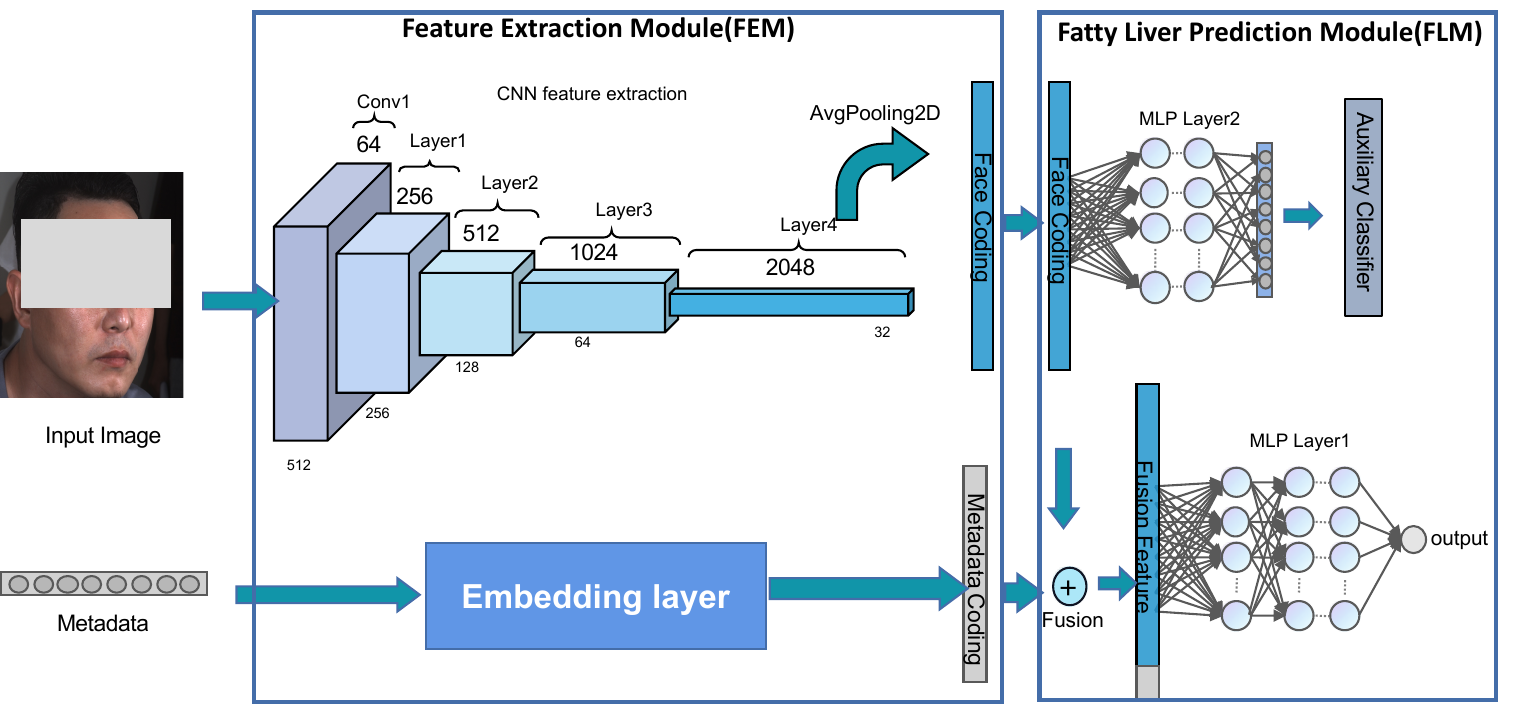}
   }
   \caption{The architecture of DeepFLD. It contains the Feature Extraction Module (FEM) and the Fatty Liver Prediction Module (FLM). First, we adopt the FEM to extract the face coding and metadata coding. The FLM is used to fuse the two codings and predict NAFLD through an auxiliary task. 
   }
   \label{fig:fusion_architecture}
   \vspace{-0.1cm}
\end{figure*}

Both metadata and face data provide important information for  NAFLD diagnosis, although face data has not been used as far as we know. In this paper, we innovatively incorporate face information for prediction and design a multi-modal fusion neural network (called DeepFLD) integrating the metadata and face data, as shown in Fig. \ref{fig:fusion_architecture}. The proposed DeepFLD model contains the Feature Extraction Module (FEM) and the Fatty Liver Prediction Module (FLM). In the FEM, the facial images  are input into a multilayer convolutional neural network to obtain the feature coding from facial images. An embedding coder layer is adopted for extracting feature coding from metadata. The prediction module FLM then combines the face and metadata coding to predict NAFLD. We also design an auxiliary task to extract effective features and accelerate  convergence during FEM training processing.

\subsubsection{Feature Extraction Module}
Generally, the facial appearance, features and expressions of patients are used by physicians to assess their health status. Theoretically, the facial image can contain all the features a physician requires to determine health status. So, we   put  the facial images into the feature extraction module. The feature extraction module is responsible for extracting high-dimensional features from the input facial images. 

A three-channel facial image ${\bf{I}}_{image} \in {\bf{R}}^{m \times n \times c_{f}}$ , is fed into multiple convolutional blocks and an average pooling layer  (${\bf{FEM}}_{CNN}$), which gradually reduces the resolution and increases the channel dimension, and finally outputs a one-dimensional feature vector ${\bf{Z}}_{image}\in {\bf{R}}^{1 \times 1 \times c_{fc}}$. During the extraction of the feature vector ${\bf{Z}}_{image}$ from facial image, it is called face coding.

The metadata is extracted from the volunteers' physical examinations and clinical laboratory assays, including age, weight, height, BMI and so on. Through the data analysis (\ref{data_analysis}), we obtain the relevant and  important metadata ${\bf{I}}_{metadata} \in {\bf{R}}^{1 \times 1 \times c_{m}}$. An embedding layer is designed to normalize each metadata element to a  N(0,1) distribution and compute the metadata coding ${\bf{Z}}_{metadata}\in {\bf{R}}^{1 \times 1 \times c_{mc}}$. Through the FEM, the inputs with different attributes in different dimensions are extracted into features with the same dimension, which is facilitated by feature fusion on the Fatty Live Prediction Module.

\[
   {\bf{Z}}_{image} = {\bf{FEM}}_{CNN}({\bf{I}}_{image}) \]
   \begin{equation}
      {\bf{Z}}_{metadata} = {\bf{FEM}}_{embedding}({\bf{I}}_{metadata})
   \end{equation}

\subsubsection{Fatty Liver Prediction Module}
The main task of the model is to predict NAFLD, so we build a multilayer perceptron (MLP) to diagnose whether the person has NAFLD.  
The whole system, especially the convolutional neural network in FEM, contains many trainable parameters. The leading result is that it may not be able to extract effective information well just by determining whether the sample has NAFLD or not. We  have also done some experiments in which we input facial images and metadata to directly predict NAFLD.  However, the results were not good enough to extract valid features related to NAFLD.
Therefore, we create an auxiliary task to help with the training of the model. The auxiliary task aims to facilitate neural network training and convergence, so some indicators related to NAFLD are considered to be selected for prediction. 
In this paper, the auxiliary task is to predict the three indicators: Gender, BMI, and Weight. 

We use MLP to build the main task network ${\bf{FLM}}_{MLP_1}$ and the auxiliary task network ${\bf{FLM}}_{MLP_2}$. Each layer of the MLP is composed of multiple neurons, and the output of the previous layer is the input of the next layer. It is also called the Fully Connected (FC) layer. In order to improve the performance of the MLP and prevent overfitting, we add a nonlinear activation layer (ReLU function) and a dropout function after each FC layer. ${\bf{FLM}}_{MLP_1}$  uses the fused data  ${\bf{Z}}_{fusion} \in {\bf{R}}^{(c_{fc}+ c_{mc}) \times 1}$ as input, which includes face  and metadata coding.  ${\bf{FLM}}_{MLP_2}$ on the other hand, only accepts face coding ${\bf{Z}}_{image}$.
\begin{equation}
   {\bf{Z}}_{fusion} = {\bf{Concat}}({\bf{Z}}_{image},{\bf{Z}}_{metadata})
   \end{equation}

We first normalize all the input elements, subtract their mean and divide their variance to make them into a positive-terminus distribution. Then, based on the fusion data ${\bf{Z}}_{fusion}$, set the number of the first channels in ${\bf{FLM}}_{MLP_1}$  and ${\bf{FLM}}_{MLP_2}$.  ${\bf{FLM}}_{MLP_1}$  has 6 FC layers, each with 2056, 1024, 1024, 512, 256, and 128 nodes.  ${\bf{FLM}}_{MLP_2}$  has three FC layers, each with 2048, 1024, and 1024 nodes.
After the fusion data is subjected to multiple nonlinear mappings in the ${\bf{FLM}}_{MLP_1}$ hidden layer, a regression value ${{y}}_{fat}$ is obtained and then normalized by the sigmoid function, and the final output is the prediction of fatty liver. Meanwhile, the fusion data is fed into ${\bf{FLM}}_{MLP_2}$, and the auxiliary task regression results ${\bf{y}}$ are output.
\[
   {{y}}_{fat} = {\bf{FLM}}_{MLP_1}({\bf{Z}}_{fusion})\]
\begin{equation}
   {\bf{y}} ={\bf{FLM}}_{MLP_2}({\bf{Z}}_{image})
\end{equation}

Convolutional neural networks trained by auxiliary tasks can also achieve NAFLD diagnosis when the metadata is removed and only images are input. However, the network without auxiliary task training cannot diagnose NAFLD very well. It shows that the designed auxiliary task is very beneficial for image training.

\subsubsection{Training}

The entire network is trained end-to-end, and the loss function is a joint loss ${\bf{L}}$, which includes the NAFLD task loss ${\bf{L}}_{NAFLD}$ as well as the auxiliary task loss ${\bf{L}}_{Auxi}$,
\[
   {\bf{L}}=\alpha{\bf{L}}_{NAFLD} +(1-\alpha){\bf{L}}_{Auxi}.
   \]
where $\alpha$ is 0.7 in this paper. The main task is  to predict whether the person has NAFLD or not, which is a binary classification task. Therefore, we adopt  classical cross-entropy. The output ${{y}}_{fat} \in [0,1]$ of module ${\bf{FLM}}_{MLP_1}$ represents the probability that the sample has NAFLD. The NAFLD task loss ${\bf{L}}_{NAFLD}$ can be expressed as follows:
\begin{equation}
   {\bf{L}}_{NAFLD} = -({\hat{y}}_{fat}log(y_{fat}) +(1-{\hat{y}}_{fat})log(1-y_{fat}))
\end{equation}
where $\hat{y}_{fat} \in \{0,1\}$ denotes the ground truth indicating whether the person has NAFLD or not. When $\hat{y}_{fat} =1 $, ${\bf{L}}_{NAFLD} = -log(y_{fat})$. Minimizing the loss ${\bf{L}}_{NAFLD}$ is equivalent to increasing $y_{fat}$,  i.e. increasing the likelihood of having a non-alcohol fatty liver. When $\hat{y}_{fat} =0 $, ${\bf{L}}_{NAFLD} = -log(1-y_{fat})$. Minimizing the loss ${\bf{L}}_{NAFLD}$ is equivalent to making $y_{fat}$ decrease, namely decreasing the probability of having NAFLD.

\section{Experiments}
For the proposed NAFLD, there are several aspects to be asked about: \textcircled{1}  the NAFLD prediction results with indicators input, and whether adding images can improve the prediction performance? \textcircled{2}  good prediction results are obtained when migrated to other data? \textcircled{3}  what are the prediction results of images as input alone?  and \textcircled{4}  whether the method prediction results have any explanations? Hence,  we perform a series of experiments.

We compare the performance of the proposed DeepFLD with multi-modal input and the models with metadata as input,  to answer the Ques. \textcircled{1}  (seen Sec. \ref{results1} ). To answer Ques. \textcircled{2}, we migrate the trained model to an unseen dataset collected in other years (seen Sec. \ref{results2}). And conducting experiments only using facial images as input to answer Ques. \textcircled{3}, (seen Sec. \ref{results3}). Moreover, ablation studies are conducted to validate 1) the significance of the clinical metadata selected by the joint indicators-based data analysis method and 2) the effectiveness of the proposed DeepFLD method with facial image input . Finally, we analyze the relationship between the face and the NAFLD through visualization to answer Ques. \textcircled{4}  (seen Sec. \ref{results4}).

\subsection{Experimental Setup}

Through data analysis, we select 8 clinical indicators that contributed to NAFLD prediction.  Since gender affects the average level of some indicators and is easily obtained, we select the top 7 indicators based on SHAP value along with gender to form 8 indicators for NAFLD prediction, denoted by  ${\bf{MetaData}}_{8}^*$ = [BMI TG HPT HLP HDL WEIGHT DRINK MALE]. And, And three non-traumatic indicators are chosen: ${\bf{MetaData}}_{3}^*$ = [BMI WEIGHT MALE]. For real-world clinical applications, we perform 5 different combinations of clinical metadata and facial images: ${\bf{MetaData}}_{8}^*$ (8 indicators), clinical metadata ${\bf{MetaData}}_{3}^*$ (3 indicators), the combination of clinical metadata ${\bf{MetaData}}_{8}^*$ (8 indicators) and facial images, the combination of clinical metadata ${\bf{MetaData}}_{3}^*$ (3 indicators) and facial images, and the facial images only.

To predict NAFLD under different types of data inputs, we built prediction methods corresponding to indicator input, image input, and joint input. We compare three models (traditional machine learning, deep machine learning, and the proposed DeepFLD) corresponding to clinical metadata as input, facial image input, and joint input. 
Random Forest (RF) and MLP are the common and effective machine learning methods for many classification tasks with one-dimensional vector input and are adopted in NAFLD \cite{R2021Assessment} and \cite{2019Prediction}. So we select them as the baseline for the NAFLD prediction under the clinical metadata as input. To fit our FLDData, we modified the methods in \cite{R2021Assessment} and \cite{2019Prediction}. In this paper, RF is a combination of 50 simple decision trees, of which the input is the medical indicators and the output is determined by the voting method. The input of  MLP is the clinical metadata, and the output is the probability of the NAFLD prediction. Specificity, we first normalize all the clinical metadata, subtract their mean and divide their variance to make them into a (0,1) distribution. Then set the number of channels in the input layer according to the input clinical metadata type (3 indicators or 8 indicators). After the input indicator is subjected to multiple nonlinear mappings in the hidden layer, a regression value is obtained and then normalized by the sigmoid function, and the final output is the prediction of NAFLD. The RF and MLP  with 8 indicators are denoted as RF (${\bf{MetaData}}_{8}^*$) and MLP (${\bf{MetaData}}_{8}^*$), and the ones with 3 indicators are denoted as RF (${\bf{MetaData}}_{3}^*$) and MLP(${\bf{MetaData}}_{3}^*$), respectively.


For the proposed model DeepFLD, we conduct experiments with three types of inputs: image only,  the combination of clinical metadata (8 indicators) and facial images, and the combination of clinical metadata (3 indicators) and facial images. Different from the CNN model, the DeepFLD model for facial image input has an auxiliary task that is to help the network learn to extract useful features. So we also conduct the experiment to compare the performance of CNN and DeepFLD with  images as the only input. Due to missing data, we select 676 samples containing images and 8 indicators for the following NAFLD prediction experiments.

\begin{figure*}[thb]
   \vspace{-0.1cm}
   \centerline{
      \includegraphics[width=0.88\linewidth]{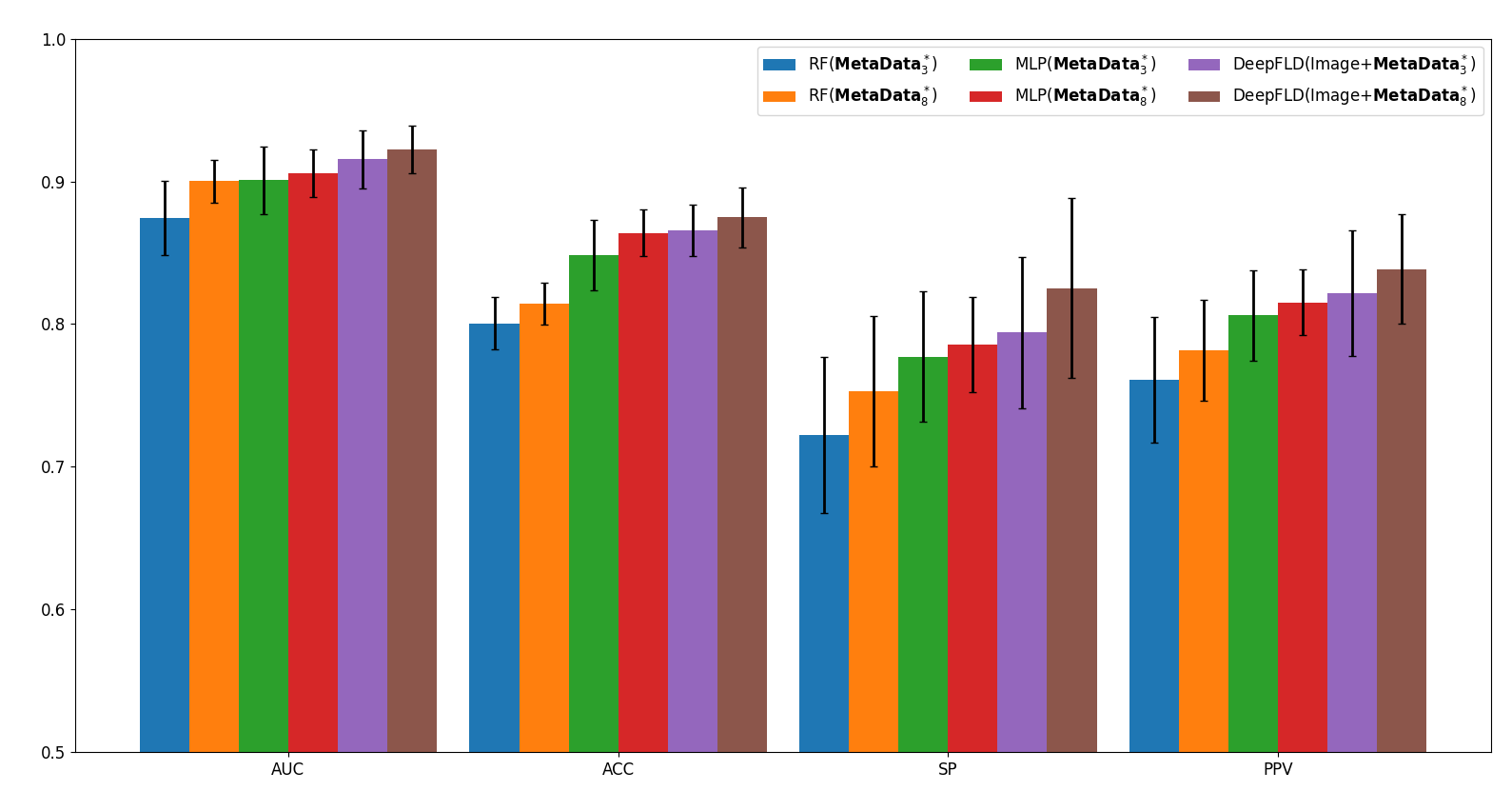}
   }
   \caption{The K-fold validation results of NAFLD prediction with different models, in which RF(${\bf{MetaData}}_{3}^*$), MLP(${\bf{MetaData}}_{3}^*$), RF(${\bf{MetaData}}_{8}^*$), and MLP(${\bf{MetaData}}_{8}^*$)  denote the Random Forest and MLP models with the selected 3 indicators ${\bf{MetaData}}_{3}^*$ and 8 indicators ${\bf{MetaData}}_{8}^*$ as input, respectively. DeepFLD(Image+${\bf{MetaData}}_{3}^*$) and DeepFLD(Image+${\bf{MetaData}}_{8}^*$)  denote the proposed DeepFLD model with the selected 3 indicators ${\bf{MetaData}}_{3}^*$ and facial images as joint input, and the  selected 8 indicators ${\bf{MetaData}}_{8}^*$ and facial images as joint input, respectively.
   }
   \label{tab_result1}
   \vspace{-0.1cm}
\end{figure*}

\subsection{Metrics}
In order to meet various screening detected or clinical applications, we take the common metrics containing classification accuracy (ACC), specificity (SP), positive predictive value (PPV),  and Area Under Curve (AUC) of different combined models for NAFLD detection.
ACC describes the classification accuracy of the classifier:
\[
ACC = \frac{(TP+TN)}{(TP+FP+FN+TN)},
\]
where TP, TN, FP, and FN denote True Positive samples, True Negative samples, False Positive samples, and False Negative samples, respectively.
SP describes the ratio of the negative samples predicted by models to all negative samples in the dataset:
\[
SP = \frac{(TN)}{(FP+TN)}   .
\]
PPV describes the ratio of the predicted true positive samples to all predicted positive samples:
\[
PPV = \frac{(TP)}{(TP+FP)} .  
\]
AUC is defined as the area enclosed by the receiver operating characteristic curve (ROC) and the coordinate axis, and its value ranges from 0 to 1. It indicates the probability that the predicted positive sample is ahead of the negative sample. For all the metrics mentioned above, the larger the value is, the better the classification effect is.

\subsection{Experimental Results}

\subsubsection{NAFLD Prediction with Multi-Modal}
\label{results1}
In this paper, we compare the NAFLD prediction results of DeepFLD with multi-modal input and the models with metadata as input. For the fairness of experiments, we adopt K-Flod cross-validation, which means the dataset is divided into K parts for cross-validation, here K=7. Each cross-validation is reiterated 7 times.  Since AUC is a comprehensive metrics to evaluate the effectiveness of models, we focus on considering this metric as the basis for model comparison.

Fig. \ref{tab_result1} shows the experimental performance, which is assessed by 7-Fold internal cross-validated metrics.
To predict NAFLD with metadata as input, we select two classical models: RF \cite{R2021Assessment} and MLP \cite{2019Prediction}, which are effectively adept at classification tasks with vector inputs. And there are 2 types of metadata as inputs: the selected 3 indicators ${\bf{MetaData}}_{3}^*$ and the selected 8 indicators ${\bf{MetaData}}_{8}^*$. Among them, ${\bf{MetaData}}_{3}^*$ contains Gender, BMI and Weight, which are easily available and non-invasive.
Their results range from  87.4\% to 91.3\%, with MLP outperforming RF.

 To obtain the NAFLD prediction results with multi-modal, including facial images and metadata, we have done 2 experiments DeepFLD (Image+${\bf{MetaData}}_{3}^*$) and DeepFLD ( Image+${\bf{MetaData}}_{8}^*$). The proposed DeepFLD achieves 92.3\% in AUC, 87.5\% in ACC metric, 82.5\% in SP and 83.9\% in PPV, which exceeds at least 1\% in AUC, 1.1\% in ACC,  5.4\% in SP, 2.4\% in PPV compared with the indicators-input models, respectively. It illustrates that facial images actually contribute to the diagnosis of NAFLD.

 Form Fig. \ref{tab_result1}, we can see that RF(${\bf{MetaData}}_{8}^*$) exceeds RF(${\bf{MetaData}}_{3}^*$)  2.6\%, MLP(${\bf{MetaData}}_{8}^*$) exceeds MLP(${\bf{MetaData}}_{3}^*$)  0.5\%  and DeepFLD (Image +${\bf{MetaData}}_{8}^*$) exceeds DeepFLD(Image+${\bf{MetaData}}_{3}^*$) 0.8\% in the comprehensive metrics AUC. It may be due that more indicators can provide more information and improve model performance. Moreover, DeepFLD(${\bf{MetaData}}_{3}^*$) also beyonds MLP(${\bf{MetaData}}_{8}^*$) in AUC, to show that NAFLD can be diagnosed only through facial image and three non-invasive indicators [Age Male BMI]. It may be due that facial images can provide more information than the indicators [TG HPT HLP HDL DRINK].

\subsubsection{Results on Unseen Dataset}
\label{results2}

\renewcommand\arraystretch{1.2} 
\begin{table}[h]
   \small
   \begin{center}
    \caption{Results of different models on new data. The second column shows the results of  cross-validations on the 2021 data. The third column shows the results of  models tested on the 2020 data, which are trained on the 2021 data.}
   \begin{threeparttable}
   \begin{tabular}{m{3.5cm}<{\centering}|p{2cm}<{\centering}|p{1.5cm}<{\centering}}
   \hline
   \hline
   Methods& AUC(2021)& AUC(2020)\\
   \hline
     	RF(${\bf{MetaData}_8^*}$) &	0.900$\pm$0.015 &0.768 			\\
   \hline
     	MLP(${\bf{MetaData}_8^*}$) &	0.906 $\pm$0.16&0.806 			\\
    \hline
    DeepFLD (Images+${\bf{MetaData}_8^*}$) & 0.923$\pm$0.17 &\bf{0.825}	\\
    \hline
    DeepFLD(Image) &0.863 $\pm$ 0.022 &0.801	\\
   \hline
   \hline
\end{tabular}
\end{threeparttable}
\label{table:unseendata}
   \end{center}
\end{table}

In order to analyze whether the model overfits the current dataset, we migrate the trained model to the data which is collected in another year.
The above results on the data collected in 2021. The trained model on the 2021 data is migrated to the data 2020 without any fine-tuning, and the results are shown in Tab. \ref{table:unseendata}. Since the datasets were collected in different years (2020 and 2021), the differences in the collection instruments lead to some differences in the distribution of indicators and facial images. Although the performance in the 2020 data is not as good as  that in the 2021 data, most models are still satisfactory, achieving 80\% in AUC. It says that these methods don't overfit the training dataset.
Moreover, the proposed DeepFLD with  multi-modal  input has the best performance with about 2\% improvement, which says it has  satisfactory generalizability.

\begin{figure*}[b]
   \vspace{-0.1cm}
   \centerline{
      \includegraphics[width=1.1\linewidth]{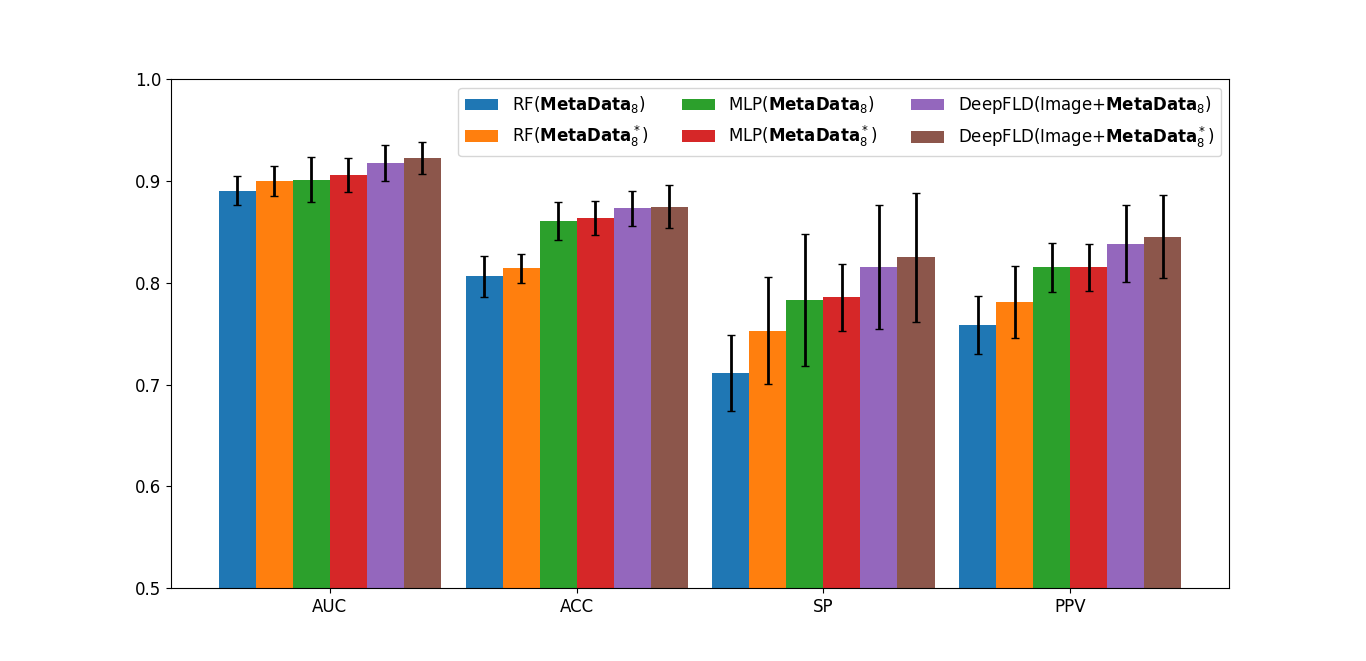}
   }
   \caption{Performance of models with different indicators, in which RF(${\bf{MetaData}}_{8}$), MLP(${\bf{MetaData}}_{8}$), RF(${\bf{MetaData}}_{8}^*$), and MLP(${\bf{MetaData}}_{8}^*$) denote Random Forest(RF) and MLP models with the input of the selected 8 indicators ${\bf{MetaData}}_{8}$ by Pearson Correlation Coefficient, and the selected 8 indicators ${\bf{MetaData}}_{8}^*$ by the joint indicators data analysis method, respectively.  DeepFLD(Image+${\bf{MetaData}}_{8}$) and DeepFLD(Image+${\bf{MetaData}}_{8}^*$) denote the proposed DeepFLD model with the selected 8 indicators ${\bf{MetaData}}_{8}$ by Pearson Correlation Coefficient and facial images as joint input, and the  selected 8 indicators ${\bf{MetaData}}_{8}^*$ by the joint indicators data analysis method and facial images as joint input, respectively.
   }
   \label{experiences2}
   \vspace{-0.1cm}
\end{figure*}
\subsubsection{NAFLD Prediction with Only Images Input}
\label{results3}
In order to obtain NAFLD prediction results with only images as input, we perform 2 experiments: cross-validation of DeepFLD  with image input only on the 2021 data, and and a migration of the trained DeepFLD model from the 2021 data to the 2020 data. The experimental results are shown in the last row of Tab.  
\ref{table:unseendata}. The experimental result of migration decrease compared to the one before migration, but is still acceptable with 80.1 \% AUC. It may be due that the brightness, tone and exposure of the images taken in different years are different.
Inspiringly,  DeepFLD with images only has competitive results compared to those with  metadata. Moreover, the result after migration (80.1\%) exceeds the one of RF model with metadata (76.8\%) by 3.3\% in AUC, paving the way for a more robust and simpler NAFLD diagnosis.

\subsection{Ablation Study}

\subsubsection{Effectiveness of the Joint Indicators-based Data Analysis}
In the Data Analysis section (Sec. \ref{data_analysis}), we adopt Pearson Correlation Coefficient and SHAP to find the most important clinical indicators for NAFLD diagnosis. So we conduct experiments with different clinical indicators. We select the top 8 indicators ranked by Pearson Coefficient Coefficient, denoted as ${\bf{MetaData}}_{8}$ =[BMI, WEIGHT, HLP, UA, TG, OBE, DBP, SBP]. We compare the performance of models under ${\bf{MetaData}}_{8}$ and ${\bf{MetaData}}_{8}^*$, seen in Fig. \ref{experiences2}. The  DeepFLD and MLP models with ${\bf{MetaData}}_{8}^*$ as input exceed the models with ${\bf{MetaData}}_{8}$ as input in  all metrics.  ${\bf{MetaData}}_{8}$ contains 6 invasive indicators and ${\bf{MetaData}}_{8}^*$ only contains 4 Invasive indicators.
It may be due that the joint indicators-based data analysis by Pearson Correlation Coefficient and SHAP can obtain the correlation of NAFLD and joint indicators. While the data analysis by Pearson Correlation Coefficient only obtains the correlation between NAFLD and the single indicator. 

\subsubsection{Effectiveness of the Medical Constraints}

\begin{figure}[htb!]
   \vspace{-0.1cm}
   \centerline{
      \includegraphics[width=0.9\linewidth]{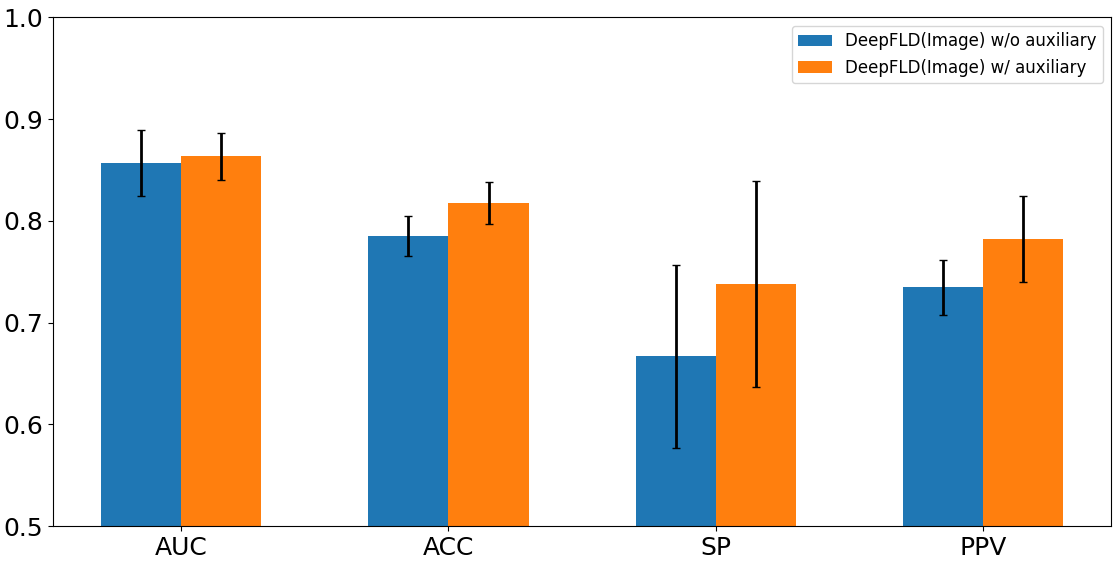}
   }
   \caption{Performance of DeepNLP without auxiliary task (DeepFLD(Image) w/o auxiliary) and DeepNLP with auxiliary task  (DeepFLD(Image) w/ auxiliary) using only facial images as input. NAFLD screening can be achieved with just a single facial image through DeepFLD, which has a high practical application value.
   }
   \label{experiences3}
   \vspace{-0.1cm}
\end{figure}

To illustrate the effectiveness of the constraints added as auxiliary tasks, we perform comparison experiments, DeepFLD(Image) without auxiliary task and DeepFLD(Image) with the auxiliary task, which are denoted as DeepFLD(Image) w/o auxiliary and DeepFLD(Image) w/ auxiliary, as shown in Fig. \ref{experiences3}. 
The poor results of DeepFLD(Image) without the auxiliary task indicate that the auxiliary task  can help the network to converge and learn the effective features to help NAFLD identification. The proposed DeepFLD achieves satisfactory diagnosis results with the designed auxiliary task. NAFLD screening can be achieved with just a single facial image through DeepFLD, which has a high practical application value.

\subsection{Visualization and Analysis}
\label{results4}
 
\begin{figure}[tbp]
   \vspace{-0.1cm}
   \centerline{
      \includegraphics[width=0.9\linewidth]{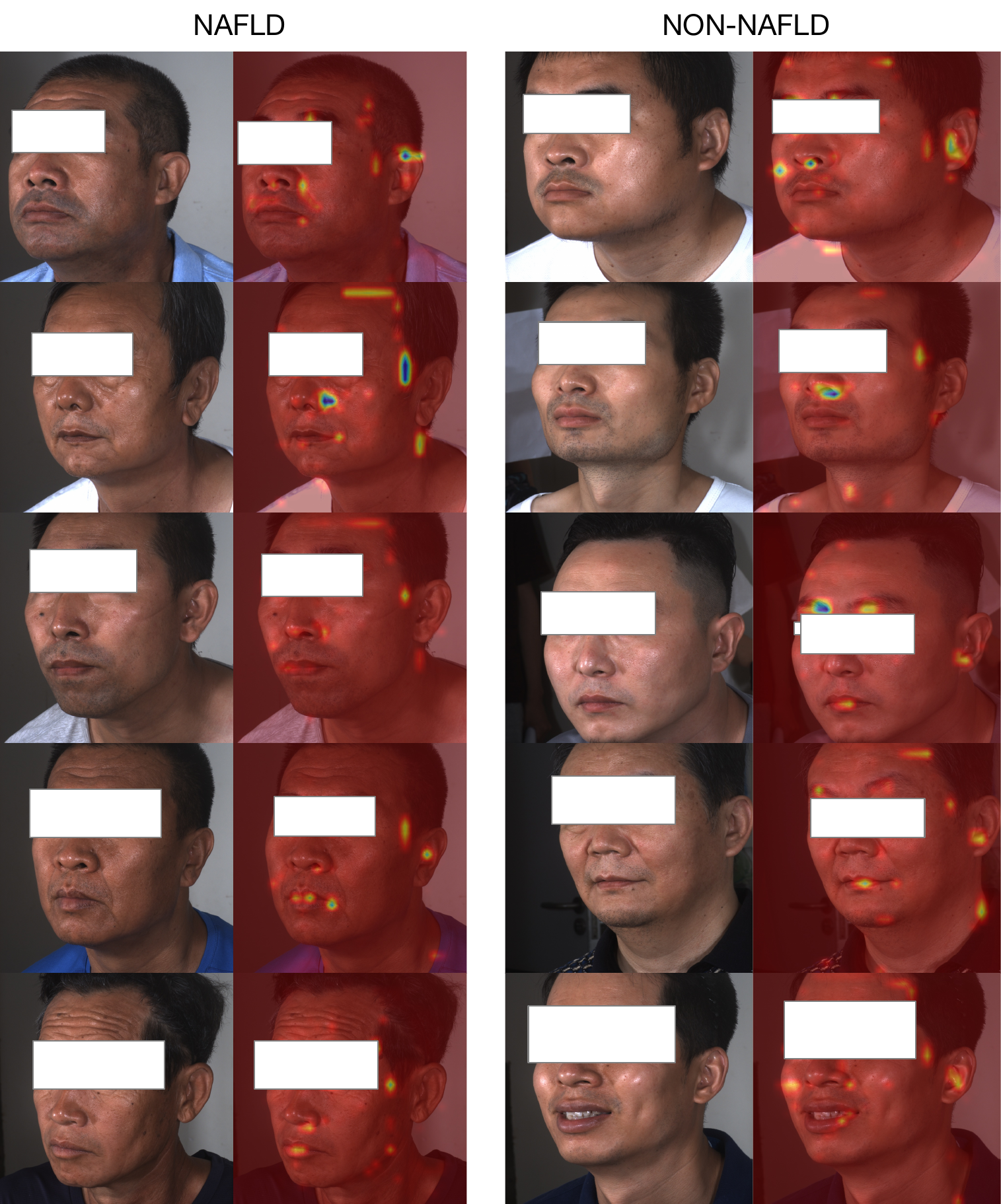}
   }
   \caption{Visualization of patients with NAFLD and healthy people (NON-NAFLD)
   }
   \label{fig:show_results}
   \vspace{-0.1cm}
\end{figure}

To explain the underlying mechanism of our study and minimize the black-box effect, the visualization technique is adopted to highlight the abnormal areas recognized by the algorithms. Fig. \ref{fig:show_results} shows the visualization of patients with NAFLD and healthy people. 

In this study, we observe that: 1) compared with healthy people, the facial skin color of patients with NAFLD is darker and yellower; and 2) patients with NAFLD have more melasma. The phenomenon of difference in skin color may be related to bilirubin\cite{RN19}. Traditionally, serum bilirubin has been used as a diagnostic marker for hepatobiliary disease, and total bilirubin elevation can occur in liver diseases\cite{RN20, RN21}. NAFLD is characterized by lipid accumulation in hepatocytes, which usually have hepatic steatosis\cite{RN23,RN22}. Antioxidative bilirubin and bilirubin-secreting biliverdin reductase  inhibit the inflammatory response of hepatocytes, intervene in the process of hepatic steatosis, and reduce liver damage\cite{RN25}. Elevated levels of bilirubin binding to the epidermis do not significantly increase skin yellowing. Moreover, bilirubin produced by cells may directly contribute to the dull appearance of facial skin\cite{RN26}. Previous studies have found that metabolic abnormalities in the liver cause hyperpigmentation and melasma\cite{RN27, RN28}. Melasma is a functional chronic disease that manifests in the face\cite{RN29}. Traditional Chinese medicine believes that the abnormal metabolism of the liver causes the metabolic waste to be effectively discharged and deposited under the skin, causing hyperpigmentation. Moreover, Fig. \ref{fig:show_results} shows that the model focuses on the facial modiolus near the mouth, which may be related to NAFLD, and needs further investigation.

\begin{figure}[tbp]
   \vspace{-0.1cm}
   \centerline{
      \includegraphics[width=1\linewidth]{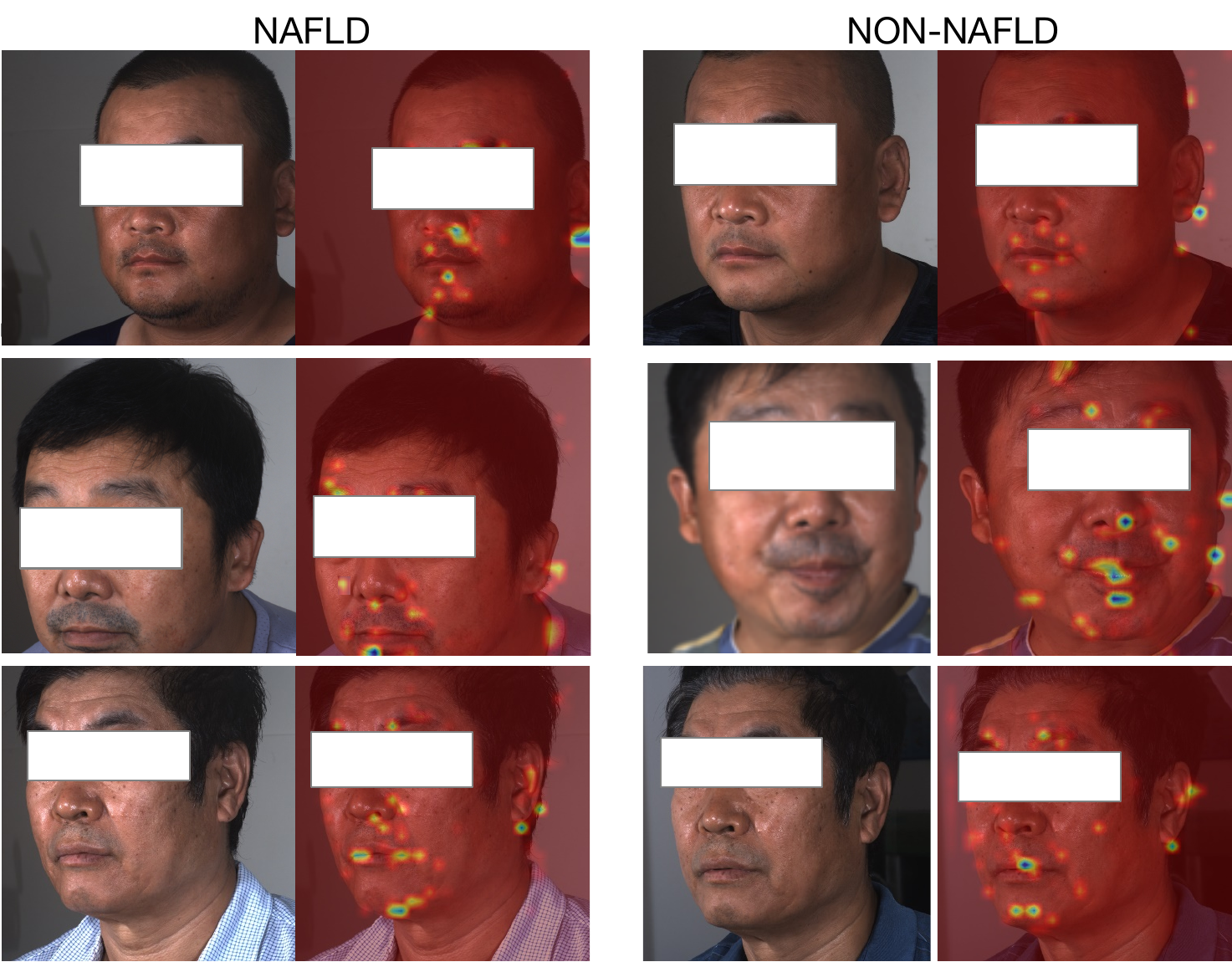}
   }
   \caption{Visualization of patients with NAFLD and healthy people (NON-NAFLD), who had a change in their liver health. Each row is the same person.
   }
   \label{fig:show_results2}
   \vspace{-0.1cm}
\end{figure}

In order to further find the differences between people with NAFLD and healthy people to explain why the model can distinguish them, we find the same people from 2020 and 2021 data, who had a change in their liver health, namely, they have  NAFLD in one year and have not NAFLD in the other year. From Fig. \ref{fig:show_results2}, we can see that skin color of the people with NAFLD is darker and yellower especially in the last row in Fig. \ref{fig:show_results2}. The second row shows the people with NAFLD have more melasma, and the last row shows the people with NAFLD is fatter. The visualization results are consistent with the findings of our analysis above,  further validating the conclusions of the above analysis.

\section{Conclusions}
This paper presents an intelligent NAFLD diagnosis system DeepFLDDiag, containing a complete clinical dataset FLDData and a novel NAFLD classification algorithm to explore whether the facial image contributes to NAFLD prediction.  FLDData includes  faces and 480 indicators of participants' physical examinations and lifestyle habits. Through joint indicators-based data analysis, we select 8 indicators which are most helpful for NAFLD classification. The proposed DeepFLD method with multi-modal input and medical constraints can 
facilitate image feature extraction from high-dimensional facial images. The proposed DeepFLD with multi-modal as input achieves  better performance than that with metadata and achieves acceptable performance on unseen data. Inspiringly, DeepFLD can further achieve competitive results using only facial images as input compared to metadata, paving the way for a more robust and simpler non-invasive NAFLD diagnosis.









\bibliographystyle{IEEEtran}
\bibliography{refs}

\end{document}